\def\etcn{$\kappa$-(BE\-DT\--TTF)$_2$\-Cu$_2$\-(CN)$_{3}$}
\def\cm{cm$^{-1}$}

\def\kcn{$\kappa$-CN}
\documentclass[aps,prl,superscriptaddress,twocolumn,showpacs,floatfix]{revtex4}
\usepackage{graphicx}
\usepackage[english]{babel}
\begin{document} %\draft
\title{Lattice vibrations of the charge-transfer salt $\kappa$-(BE\-DT\--TTF)$_2$\-Cu$_2$(CN)$_{3}$: \\
novel interpretation of the electrodynamic response in a spin-liquid compound}
\author{M.\ Dressel}
\affiliation{1.~Physikalisches Institut, Universit\"{a}t
Stuttgart, Pfaffenwaldring 57, 70550 Stuttgart Germany}
\author{P. Lazi\'{c}}
\affiliation{Rudjer Bo\v{s}kovi\'{c} Institute,
Bijeni\v{c}ka cesta 54, HR-10000
Zagreb, Croatia}
\author{A.\ Pustogow}
\affiliation{1.~Physikalisches Institut, Universit\"{a}t
Stuttgart, Pfaffenwaldring 57, 70550 Stuttgart Germany}
\author{E.~Zhukova}
\affiliation{1.~Physikalisches Institut, Universit\"{a}t
Stuttgart, Pfaffenwaldring 57, 70550 Stuttgart Germany}
\affiliation{A.M. Prokhorov General Physics Institute, Russian Academy of Sciences,
%Vavilov str.\ 38,
119991 Moscow, Russia}
\affiliation{Moscow Institute of Physics and Technology (State University), 141700, Dolgoprudny, Moscow Region, Russia}
\author{B.~Gorshunov}
\affiliation{1.~Physikalisches Institut, Universit\"{a}t
Stuttgart, Pfaffenwaldring 57, 70550 Stuttgart Germany}
\affiliation{A.M. Prokhorov General Physics Institute, Russian Academy of Sciences,
%Vavilov str.\ 38,
119991 Moscow, Russia}
\affiliation{Moscow Institute of Physics and Technology (State University), 141700, Dolgoprudny, Moscow Region, Russia}
\author{J.\ A.\ Schlueter}
\affiliation{Material Science Division, Argonne National Laboratory - Argonne, Illinois 60439-4831, U.S.A.}
\author{O. Milat}
\affiliation{Institut za fiziku, P.O.Box 304, HR-10001 Zagreb, Croatia}
\author{B. Gumhalter}
\affiliation{Institut za fiziku, P.O.Box 304, HR-10001 Zagreb, Croatia}
\author{S.~Tomi\'{c}}
\affiliation{Institut za fiziku, P.O.Box 304, HR-10001 Zagreb, Croatia}
\date{\today}

\begin{abstract}
The dimer Mott insulator
$\kappa$-(BEDT-TTF)$_2$Cu$_2$(CN)$_3$ exhibits unusual electrodynamic properties.
Numerical investigations of the electronic ground state and the molecular and lattice vibrations reveal the importance of the Cu$_2$(CN)$_3^-$ anion network coupled to the BEDT-TTF molecules: The threefold cyanide coordination of copper and linkage isomerism in the anion structure cause a loss of symmetry, frustration, disorder, and domain formation.
Our findings consistently explain the temperature and po\-lar\-i\-zation-dependent THz and infrared measurements,
reinforce the understanding of dielectric properties and have important implications for the quantum spin-liquid state, which should be treated beyond two-dimensional, purely electronic models.
\end{abstract}


\maketitle

%\section{Introduction}

In the field of quasi-two-dimensional strongly correlated electron systems, the $\kappa$-salts
of the BEDT-TTF family have attracted considerable interest \cite{Kanoda11}
because their triangular arrangement of the dimers is close to geometrical frustration
leading to the first realization of spin-liquid behavior
in \etcn\ \cite{Shimizu03}. The antiferromagnetic exchange coupling among the dimers is rather strong ($J\approx 250$~K), however, neither the magnetic order nor structural distortions
have been detected down to lowest temperatures.
Despite enormous theoretical and experimental efforts, the nature of this phase is still inconclusive and the question of a possible gap in the spin excitations remains open \cite{Yamashita08,Yamashita09,Kaneko14,Lee08}.
From the residual spin susceptibility and the power-law temperature dependence of
the NMR spin-lattice relaxation rate,
low-lying spin excitations are concluded, while their contribution to the optical conductivity is still under discussion
\cite{Ng07,Kezsmarki06,Elsaesser12}.

\begin{figure}[b]
 \centering
 \includegraphics[width=1\columnwidth]{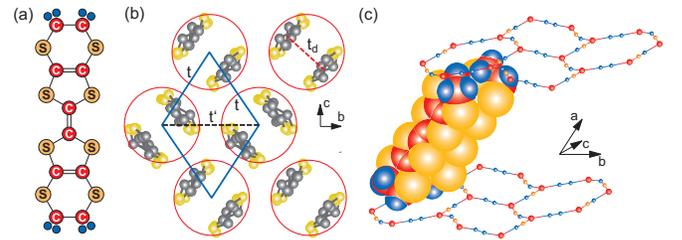}
\caption{\label{fig:1} (Color online) (a)~Sketch of the
bis-(ethyl\-ene\-di\-thio)\-te\-tra\-thia\-ful\-va\-lene
molecule, called BEDT-TTF. (b)~For $\kappa$-(BEDT-TTF)$_2$Cu$_2$(CN)$_3$ they
are arranged in dimers, which constitute an anisotropic triangular lattice
within the $bc$ conduction layers.
(c)~These BEDT-TTF layers are sandwiched along the $a$-axis by insulating anion sheets.
The polymerized Cu$_2$(CN)$_3^-$ form  interconnected elongated
hexagons with the terminal ethylene groups of a BEDT-TTF dimer fitting into the opening.
The ratio of the inter-dimer transfer integrals is close to unity:
$t^{\prime}/t \approx 0.83$ \cite{Nakamura09,Kandpal09}; the intra-dimer transfer integral $t_d\approx 0.2$~eV \cite{Oshima88,Komatsu96}.
With $U\approx 2t_d$ \cite{McKenzie98}, one obtains at ambient
conditions $U/t=7.3$; a slight variation of $U$ and orbital overlap $t$
causes magnetic order, on the one hand, and superconductivity, on the other hand.
}
\label{fig:structure1}
\end{figure}

Interacting spins on a triangular lattice cannot order in a simple antiferromagnetic ground state,
quantum fluctuations prevent a stable spin-liquid phase in the presence of finite hopping $t$ and electronic repulsion $U$.
More advanced microscopic models have been put forward to explain the properties of $\kappa$-BEDT-TTF salts \cite{Hotta10,Naka10,Mazumdar10,Gomi13} based on the presence of charge dipoles on the dimers which couple to spin degrees of freedom and thereby prevent ordering.
While the electronic order has been proven in related compounds \cite{Drichko14} in which the intra-dimer Coulomb repulsion $U$ is reduced with respect to inter-site interaction $V$, the ratio $V/t$ is of minor relevance in the case of \etcn, called \kcn\ (Fig.~\ref{fig:structure1}). The anomalous dielectric response observed in the kHz and MHz range below 60~K \cite{Abdel-Jawad10,Pinteric14} evidences some unexpected charge excitations in the spin-liquid state which must be explained satisfactorily.
Recently, time-domain optical measurements rendered a rather broad absorption band around 1~THz that develops below $T\approx 80$~K. Itoh {\it et al.}  \cite{Itoh13} attributed it to collective excitation of intra-dimer electric dipoles. However, NMR, Raman and infrared spectroscopies rule out any sizable charge imbalance all the way down to low temperatures \cite{Shimizu06,Yamamoto12,Sedlmeier12}.

In order to shed light on the excitation properties in the quantum spin-liquid state in general,
and the prototypical dimer system \kcn\ in particular,
we have performed broad-band optical investigations
with the emphasis on the THz and far-infrared spectral range complemented by the first extensive DFT calculations of the electronic structure and vibrational properties. Thereby we bridge the gap between microscopic models and macroscopic observations.
We find that in \kcn\ not only the triangular arrangement of the dimers is important, but also the anions exhibit intrinsic frustration and disorder due to polar cyanide links between the three-fold bound Cu ions. Specifically, we make clear that
the low-energy modes in the THz range are lattice vibrations rather than collective dimer
excitations: the dominant motion of the Cu$_2$(CN)$_3^-$ anions is strongly coupled to the BEDT-TTF dimers.

\begin{figure}[b]
\centering
\includegraphics[clip=true,width=0.63\columnwidth]{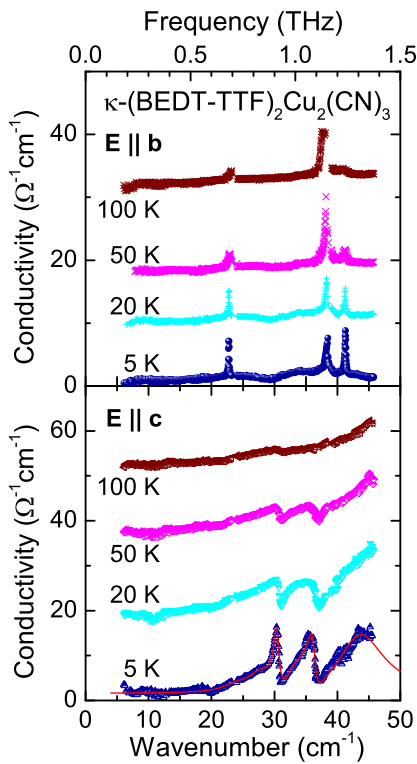}
\caption{(Color online) THz-spectra of the optical conductivity of \etcn\ measured at different temperatures for both polarizations using a coherent-source Mach-Zehnder interferometer.
The data are shifted with respect to each other for clarity reasons.
The red line corresponds to the fit of the data.}
\label{fig:THz-mode}
\end{figure}

%\section{Experimental Details}
The in- and out-of-plane infrared reflectivity
of \kcn\ single crystals was measured with 1~\cm\ spectral resolution
using several Fourier-transform spectrometers
equipped with proper polarizers in order to probe the electrodynamic response in different directions and at the temperatures $300~{\rm K}>T>5$~K.
The optical conductivity is obtained by standard Kramers-Kronig analysis; for further detail see Refs.~[\onlinecite{Elsaesser12,Sedlmeier12}]. At very low frequencies $\nu < 45$~\cm\ \^{=} 1.5~THz, we were able to perform transmission experiments on a very large single crystal ($2\times 4$~mm$^2$) as for $T\leq 100$~K the absorption becomes small enough.
Employing a coherent source THz Mach-Zehnder interferometer \cite{Gorshunov05}
both the amplitude and phase change can be  measured independently with high spectral resolution and accuracy. The optical conductivity %and permittivity are
is directly calculated by Fresnel's equations
and plotted in Fig.~\ref{fig:THz-mode} for different polarizations and temperatures as indicated. The corresponding permittivity is shown in Fig.~S1 of the Supplemental Materials \cite{SM}.

%\section{Experimental Results}
For $E\parallel b$ we observe a phonon-like doublet around 40~\cm\
and another mode around 23~\cm.
As expected for a vibrational feature, the latter one continuously
narrows and becomes more pronounced as the crystal is cooled.
Interestingly, when the temperature is reduced below 60~K a rather sharp peak at 41.3~\cm\ grows drastically at the expense of the 38.5~\cm\ mode.
Also for $E\parallel c$ between 30 and 40~\cm\ actually two distinct features
become very prominent at low temperatures.
In accord with Ref.~\cite{Itoh13}, the peaks grow for $T<60$~K;
they have a rather unusual sawtooth shape, that can only be fitted satisfactorily by two Fano functions and a Lorentzian on top of a power-law background \cite{Elsaesser12}.

%\section{Calculations}
In order to assign and finally understand the observed vibrational features,
we have carried out first-principles calculations of the electronic structure in the framework of the density functional theory (DFT) as implemented in the VASP code using the projector augmented-wave method \cite{Blochl94,Kresse93,SM}.
%\section{Discussion}
The structural data obtained by x-ray investigations at $T=100$~K are solved in $P2_1$  symmetry and reduced to $P1$ which turns out to be the relaxed structure of lowest energy.
The obtained electronic band structure of the \kcn\ system
is in accord with previous calculations \cite{Nakamura09,Kandpal09} and plotted in Fig.~S3;
two antibonding bands of BEDT-TTF molecules cross the Fermi level and the bonding bands lie well below it.  The overall cation-derived character of kappa-CN band structure around the Fermi level is also confirmed by the rigid band shift of molecular subsystem alone that is induced by the charge transfer of nearly two electrons \cite{SM}. Despite the cation-band character around the Fermi level, the anion layer turns out to be crucial for the full understanding of this charge transfer salt.

\begin{figure}
\centering
\includegraphics[clip=true,width=0.6\columnwidth]{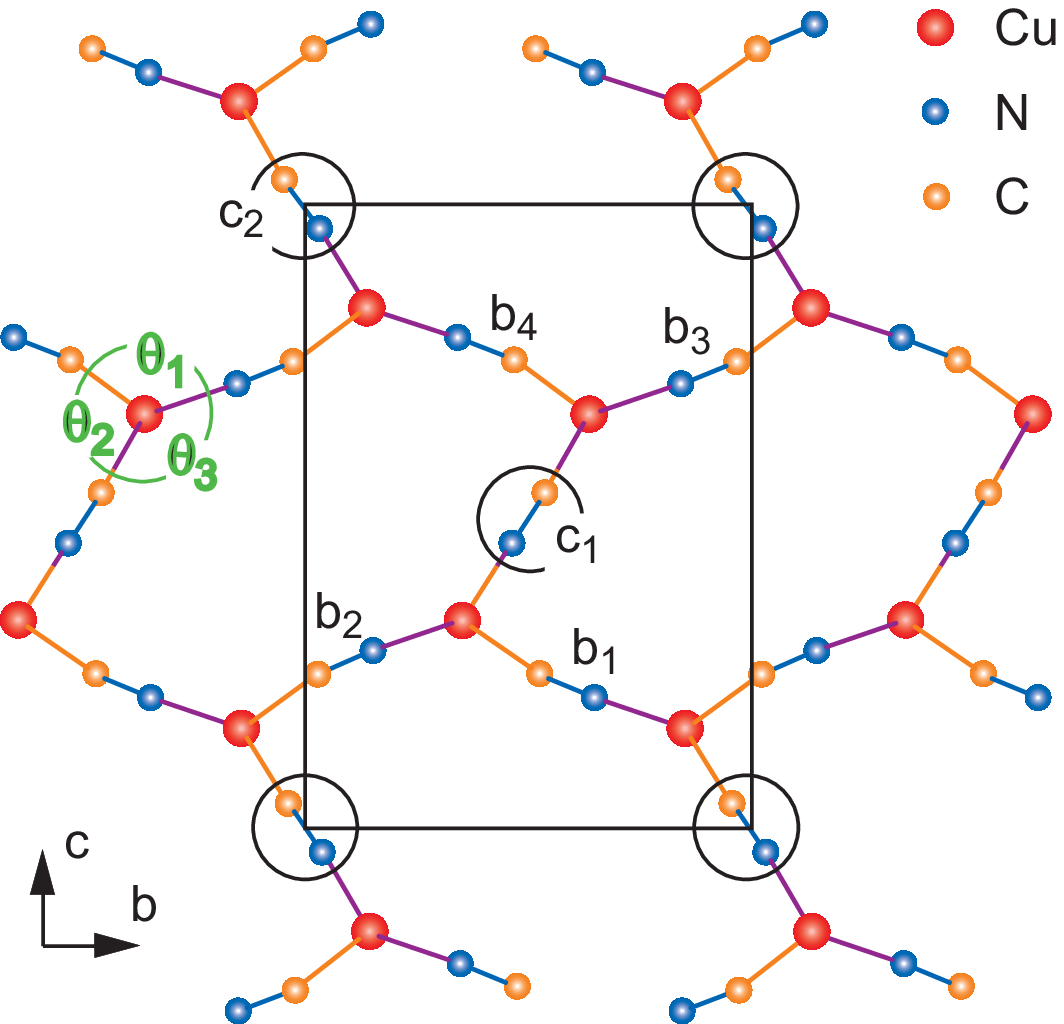}
\caption{(Color online) Relaxed structure of the Cu$_2$(CN)$_3^-$ anion layer in the
$bc$ plane of \etcn\  as obtained by the DFT calculations. The unit cell contains two formula units ($Z=2$) with two distinct configurations of Cu ions. When
the positions of C and N in the encircled links are specified, the symmetry
is reduced to P1. For the sake of simplicity, only the inequivalency between the bridging c$_1$ and c$_2$ is shown. The coordination at the Cu site ($\theta_1=118.7^{\circ}$, $\theta_2=102.3^{\circ}$, $\theta_3=138.8^{\circ}$) is closer to the ideal $120^{\circ}$ than for other polyanions \cite{Hiramatsu15}.
This constellation is prone to frustration, disorder and domain formation.}
\label{fig:structure2}
\end{figure}

Hence we now turn to the Cu$_2$(CN)$_3^-$ network; in Fig.~\ref{fig:structure2} we can identify two segregated chains mainly directed along the $b$-axis.
Each Cu is linked to the three neighboring Cu ions by cyanide groups; in one of them the CN are arranged in positive direction,
here labeled as b$_1$ and b$_2$, while the groups b$_3$ and b$_4$ point in the negative direction. Inverting one CN costs 174~meV; if all four groups are oriented parallel, the energy of the lattice is enhanced by 240~meV.
This falls into the energy range of few hundred meV known for a cyanide flip in the case of linkage isomerism \cite{Bignozzi94,Shatruk07,Dumont13}.

Even more interesting are the two bridging links c$_1$ and c$_2$ for which the cyano groups are directed dominantly along the $c$-axis. Commonly the structure is solved within the $P2_1/c$ symmetry implying that due to the linkage isomerization they sit at inversion centers, i.e. the orientation is stochastic \cite{Geiser91}.
Recent x-ray scattering studies \cite{Pinteric14} already suggested that the lower symmetry group may be more relevant;
this is now confirmed by the present calculations as well as by the targeted structural studies \cite{Foury2015}:
the crystallographic structure of \kcn\ is best characterized by the low symmetry group $P1$.
The energy minimum is reached for both c-cyanides aligned in the same direction as pictured in Fig.~\ref{fig:structure2}. However, the
energy difference relative to the situations of reversed (parallel as well as antiparallel) cyanides is only 10 to 15~meV, i.e. one order of magnitude less than for the inversion of cyanides in the $b$-chains.
From the Cu-ion point of view, the linkage isomerism causes frustration as they may be coordinated to two C and one N atoms, or to one C and two N, paving the way for intrinsic disorder. The consequence is a local rearrangement of the lattice. Naturally, the domains of local order of one or the other preferential direction will form. Coupling via hydrogen bonds to the ethylene groups extends the distortions of the anion network onto the BEDT-TTF layers.
This fact might explain the relaxor-like low-frequency dielectric response observed previously \cite{Abdel-Jawad10,Pinteric14}.

\begin{figure}
\centering
\includegraphics[clip=true,width=0.9\columnwidth]{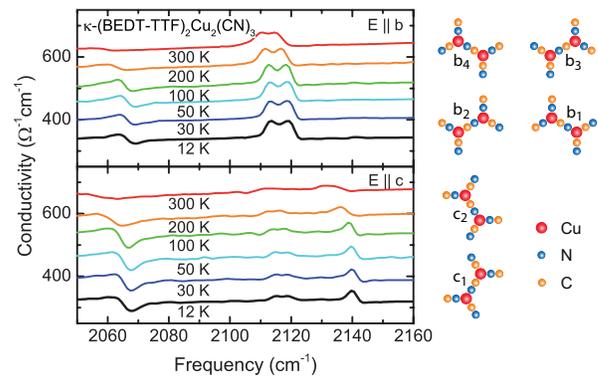}
\caption{(Color online) Mid-infrared conductivity of \etcn\ recorded for
the polarization $E\parallel b$ and $E\parallel c$ at different
temperatures. The CN stretching modes are seen between 2115 and 2140~\cm.
On the right the different cyanide arrangements in the lowest energy state are sketched.
For $E\parallel c$ the two arrangements c$_1$ and c$_2$ are degenerate in energy.
}
\label{fig:mir-mode}
\end{figure}

The phonon modes are calculated with the PHONOPY code \cite{Togo08} using the DFT-derived force constants \cite{Kresse93}.
Among the 360 vibrations obtained in the spectral range
up to approximately $3000~{\rm cm}^{-1}~\hat{=}~90$~THz \cite{SM}
let us consider first the stretching motion of the anions
where the cyanides move between the heavy Cu ions basically without affecting the other constituents.
The 2139~\cm\ feature observed for $E\parallel c$ (Fig.~\ref{fig:mir-mode}) can be assigned to
vibrations of the c$_1$ and c$_2$ groups calculated to appear at $\nu_{\rm 327}=2131.8$~\cm\ and $\nu_{\rm 328} = 2136.0$~\cm,
for the out-of-phase and in-phase motion, respectively.
The four b-type CN groups are expected to oscillate at 2089.9,  2095.1,
2101.6, and 2106.8~\cm\ and can be associated with the double structure seen
in Fig.~\ref{fig:mir-mode} at 2113 and 2118~\cm\ for $T=100$~K, as an example.
Other cyanide modes are identified at around 320 and 500~\cm\ (cf.\ Fig.~S2).
The important result from our infrared data is a rather strong anisotropy of the cyanide vibrational features. The bridging c$_1$and c$_2$ groups possess larger force constants than the four b-type CN groups. This may be due to shorter and more rigid Cu-CN-Cu bonds, implying steeper potentials for the bridging than for the four b-type CN groups. Such a contribution prevails over the standard charge-density term. Namely, our calculations of the cation-anion interaction-induced charge distribution demonstrate that at the cyanide sites the electron density is enhanced compared to the bare one without the BEDT-TTF molecules; this effect is much stronger along the $b$-chains than along the bridging $c$-axis (Figs.~S4 and S5).

More remarkable, however, are the pronounced modes in the THz range, which involve bending and twisting motions of larger entities, or of the anions as a whole. Our calculations evidence that these low-energy features contain the anions as well as the BEDT-TTF cations, i.e.\ they are  lattice phonons rather than molecular vibrations \cite{Dressel92,Girlando00,Dressel12}. The cyano groups in \kcn\ possess static dipole moments and interact with BEDT-TTF molecules; these polar links are found to be larger for the four b-type cyanides than for the bridging c$_1$ and c$_2$-bonds (Fig.~S5).

\begin{table}
\caption{Vibrational modes of \etcn\ in the THz range. The observed features of Fig.~\ref{fig:THz-mode} are compared to modes obtained by {\it ab-initio} phonon calculations based on density-functional-theory. A complete list and 3D motion pictures are given in the Supplemental Materials \cite{SM}.\\
\label{Tab:1}}
%\begin{center}
\begin{tabular}{c|cc|cc}
%\hline
Mode & \multicolumn{2}{c|}{$E\parallel b$} & \multicolumn{2}{c}{$E \parallel c$}  \\
No.  & calc.     & exp.                      & calc & exp.         \\
  & ~(cm$^{-1}$)~  & ~(cm$^{-1}$)~               & ~(cm$^{-1}$)~ & ~(cm$^{-1}$)~  \\
  \hline
4    &   24.3    &   22.7                    &       &       \\
5    &           &                           & 33.3  & 31.2  \\
8    &           &                           & 37.8  & 37.0  \\
9    &    38.2   &   38.4                    &       &       \\
11   &    40.5   &   41.3                    &       &
\vspace*{-3mm}
%\hline
\end{tabular}
%\end{center}
\end{table}
The computed frequencies of the long-wavelength vibrations are in good agreement with the measured optically excited phonons in the THz range (see Fig.~\ref{fig:THz-mode}), as demonstrated in Tab.~\ref{Tab:1}.
For $E\parallel b$ two modes are calculated to appear at $\nu_{9}=38.2$ and at $\nu_{11}=40.5$~\cm,
which involve coupled cation-anion motion: bending of the complete dimer followed by shear and waving motion of the anion layer. The outer rings of BEDT-TTF are particularly affected by the higher-frequency excitation.
The corresponding features for $E\parallel c$
are related  to twist motions of the BEDT-TTF dimer with a strong involvement
of the ethylene groups; again the dimer motion is coupled to shear motion of the anion layer.

The phonon lineshape for the two polarizations shown in Fig.~\ref{fig:THz-mode}
is notably different:  in the case of $E\parallel b$
they exhibit a simple Lorentzian structure, whereas for $E\parallel c$
these are asymmetric Fano-like lineshapes.
In the first case the photons couple directly to the optical phonons \cite{remark1};
the effect becomes even stronger as the charge density in the anion layers is enhanced along the $b$-direction
by the cation-anion interaction.
For the $c$-polarization, additional resonant interactions with a continuum of excitations must be invoked to
cause Fano-interference effects.
Spinons have been suggested as candidates for these low-energy excitations responsible for the concomitant increase of the background THz conductivity \cite{Ng07,Kezsmarki06}. However, the experimentally observed exponents differ by more than a factor of two from the theoretically predicted ones \cite{Elsaesser12}.
Alternatively, the excitation continuum in the THz range may be associated with intraband electronic excitations, i.e.\ the creation of soft electron-hole pairs across the Fermi level in the $\Gamma$Z and $\Gamma$Y directions of the first Brillouin zone (cf.\ Fig.~S3). However, photons cannot induce
these electronic transitions directly because of momentum conservation.
The mechanism of resonant excitation of electron-hole pairs by optical phonons \cite{Kuzmenko} requires
that one photon simultaneously creates an intraband electron-hole pair and a sub-THz phonon with opposite momentum in the end-vertices of the optical response function.
Such interactions are discussed in detail in Ref.~\cite{remark1}. Together with the selection rules favoring the photon coupling to intraband excitations for $E\parallel c$,
this excitation scenario can explain the different lineshapes seen in Fig.~\ref{fig:THz-mode}.
The intensity of the phonon peaks diminishes as the temperature increases.
This may indicate that the phase space for the excitation of soft electron-hole pairs across the Fermi level becomes blurred; the smearing of sharp phonon features in electron-phonon interactions may also be caused by enhanced anharmonic couplings.

Our findings demonstrate that the strong THz features cannot be caused by
collective excitations of the electric dipoles fluctuating within the dimers as suggested in Ref. \cite{Itoh13}. Rather, they are due to the coupled BEDT-TTF-anion vibrations involving the entire lattice.
The relation to the low-frequency dielectric response that occurs below $T\approx 60$~K \cite{Abdel-Jawad10} is certainly not by microscopic dipoles within the BEDT-TTF dimers
which have been ruled out by vibrational spectroscopy \cite{Sedlmeier12}.
Instead we suggest that the relaxor ferroelectric mode is related to the charge inhomogeneities caused by frustration and disorder in the triangular pattern of the Cu$_2$(CN)$_3^-$ anion layers originating in
the isomorphism in the cyanides which carry a rather strong dipole moment.
The coupling of organic molecules to the anion layers via the ethylene groups imposes large-scale charge inhomogeneities and domain walls onto the BEDT-TTF layer. This is in fact what is observed in the macroscopic dielectric response \cite{Abdel-Jawad10,Pinteric14}. Our conclusions are in line with calculations \cite{Pouget} that show the essential role of anions at the charge ordering transition of $\alpha$-(BEDT-TTF)$_2$I$_3$.
In the present case, however, no static charge-order develops but two highly-frustrated subsystems interact.

%\section{Conclusion}

Our comprehensive experimental and numerical investigations of \etcn~
combine optical measurements with the electronic structure and vibrational dynamics calculations. They enable us to understand how the intrinsic Cu$_2$(CN)$_3^-$ anion structure causes frustration, disorder, loss of symmetry and domain formation.
Thereby we can consistently explain
the absence of charge disproportionation on the BEDT-TTF dimers,
the signatures of cyanide vibrations, the pronounced anisotropic phonon modes in THz regime, and the relaxor-like dielectric response occurring below $T=60$~K.
The intrinsic domain structure and frustration within the coupled BEDT-TTF and Cu$_2$(CN)$_3^-$ anion system
are the origin of the electrodynamic response. This immediately implies that the interaction with the anion layer and the structure of Cu$_2$(CN)$^-_3$ network is crucial for understanding the quantum spin-liquid state in \etcn. Further experimental and theoretical studies will clarify whether this new level of complexity
provides the missing link to stabilize this unusual ground state.

\begin{acknowledgements}
We would like to thank P. Foury, V. Ilakovac, \text{J.-P.} Pouget and G. Saito for many enlightening discussions.
We acknowledge financial support by the Deutsche Forschungsgemeinschaft (DFG), Deutscher Akademi\-scher Austauschdienst (DAAD), the Russian Ministry of Education and Science (Program 5 top 100) and the Croatian Science Foundation project IP-2013-11-1011.
\end{acknowledgements}

\newpage

\begin{center}
{\LARGE Supplemental Material}
\end{center}

\renewcommand{\baselinestretch}{1.2}

\section{THz response}
\begin{figure}[h]
\centering
\renewcommand{\thefigure}{S\arabic{figure}}
\includegraphics[clip=true,width=0.8\columnwidth]{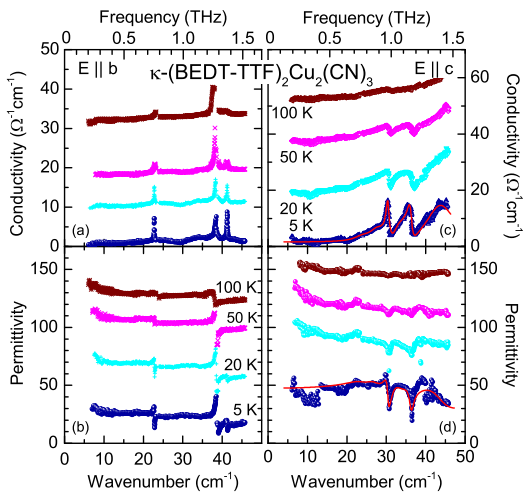}
\caption{Frequency dependence of (a,c) the optical conductivity and (b,c) the dielectric constant of \etcn\ measured in the THz range at different temperatures for both polarizations (a,b) $E\parallel b$ and (c,d) $E\parallel c$
using a coherent source Mach-Zehnder interferometer.
The data are shifted with respect to each other for clarity reasons.
The red line corresponds to the fit of the data by two Fano functions and a Lorentzian on top of a power-law background.}
\label{fig:S1}
\end{figure}

Using a Mach-Zehnder interferometer in the frequency domain allows us to measure independently the optical transmission and phase change in the THz frequency range from 6 to 46~\cm\ with an extremely high spectral resolution of 0.05~\cm. The frequency dependent conductivity $\sigma(\nu)$ and permittivity $\epsilon(\nu)$ are then calculated by using Fresnel's equations for each frequency \cite{DresselGruner02,Gorshunov05} without assuming any model.
In Fig.~\ref{fig:S1} the optical conductivity (reproduced from Fig.~2 of the main text) is complemented by the permittivity for both polarizations $E\parallel b$ and $E\parallel c$ at different temperatures.
It is interesting to note the pronounced anisotropy in the dielectric constant: for the lowest frequency we find $\epsilon_b \approx 30$ while $\epsilon_c \approx 50$ in the $c$-direction is more or less independent on temperature. This observation is in accord with the anisotropic response in the conductivity.

\section{Infrared vibrations}
\begin{figure}[h]
\renewcommand{\thefigure}{S\arabic{figure}}
\centering
\includegraphics[clip=true,width=0.8\columnwidth]{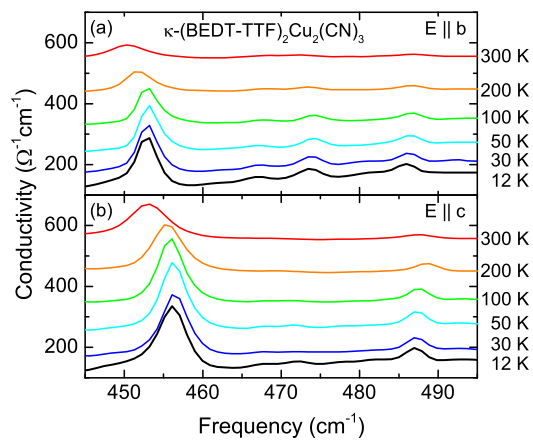}
\caption{Far-infrared conductivity spectra of \etcn\ recorded for
the polarization $E\parallel b$ and $E\parallel c$ at different temperatures as indicated.
}
\label{fig:FIR-mode}
\end{figure}

Besides the THz lattice vibrations displayed in Fig.~\ref{fig:S1} and the typical
molecular vibrations in fingerprint mid-infrared regime, plotted in Fig.~4 of the main paper, we do also observe
characteristic features of the cyanide vibrations in the far-infrared spectral range, as displayed in Fig.~\ref{fig:FIR-mode}. These features are associated with vibrational motions of BEDT-TTF molecules with a strong participation of the ethylene groups. The mode seen at 487~\cm\ for $E\parallel c$ is related to a vibrational motion of the cyanide groups with respect to the copper ions.
For $E\parallel b$ two separated features are observed at 473 and 486~\cm, which are related mainly to the b$_2$ and b$_4$ bonds in one case and b$_3$, b$_4$ and b$_1$, b$_2$ in the other case.

It is interesting to note that for both polarizations the 455~\cm\ modes shift to higher frequencies when the temperature is reduced from $T=300$~K to 50~K, but seem to red-shift slightly at lower temperatures. The 487~\cm\ features do not exhibit this hardening at elevated temperatures. More detailed studies with higher resolution are in progress.

%\newpage
\renewcommand{\baselinestretch}{1.2}
\section{Computational details, electronic structure and vibrational frequencies}
We have employed the self-consistently implemented
van der Waals density functional (vdW-DF) \cite{Dion04,Roman09,Klimes11}
for correlation with optB88 for exchange \cite{Mittendorfer11}.
The expansion in plane waves was done with a cutoff energy of 700~eV,
the Brillouin zone was sampled by $1 \times 2 \times 2$ Monkhorst-Pack choice of $k$-points \cite{Monkhorst76}.
We relaxed the full unit-cell shape/volume and the atomic positions
til the forces on the atoms dropped below 1~meV/\AA.

\begin{figure}[h]
\centering
\renewcommand{\thefigure}{S\arabic{figure}}
\includegraphics[clip=true,width=1\columnwidth]{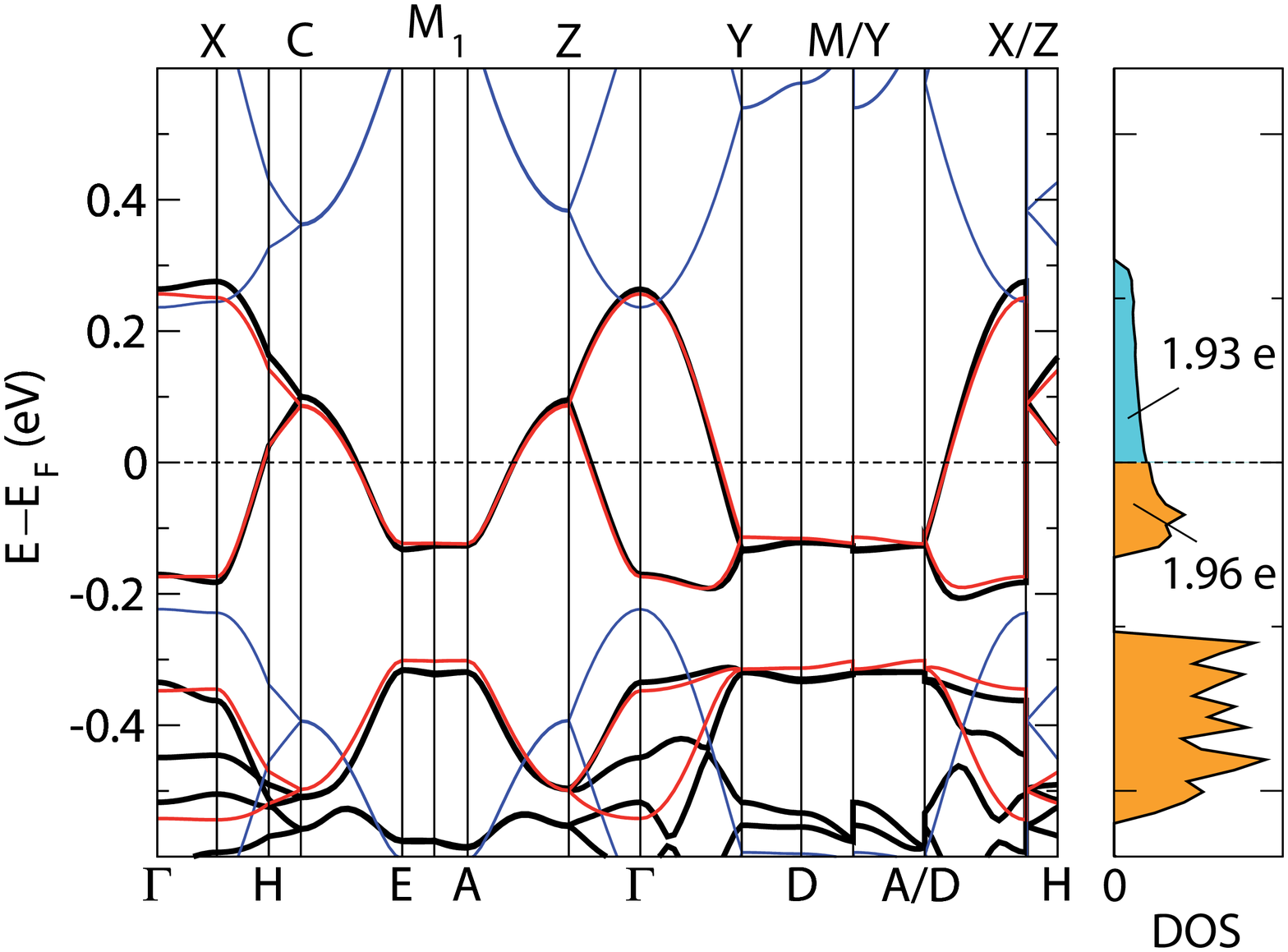}
\caption{Band structure of \etcn\ (black) plotted along the high-symmetry directions. The segments $\Gamma Z$ and $\Gamma Y$ in the first Brillouin zone correspond to the $c$- and $b$-directions, respectively.
If only the molecular subsystem is considered, the generic cation bands (blue) are obtained.
Shifting them by 480~meV (red) results in an almost perfect coincidence with
the complete band structure of \etcn,  pointing to the overall cation-derived character around the Fermi level.
The occupation of the corresponding density of states (right panel), indicates a half-filled band, in agreement with the chemical composition of dimerized systems.
}
\label{fig:band structure}
\end{figure}

The electronic band structure of the whole \etcn\ system is plotted by black lines in Fig.~\ref{fig:band structure};
two antibonding bands of BEDT-TTF molecules cross the Fermi level and the bonding bands lie well below it. Most interestingly, the generic cation bands (blue lines) calculated solely for the molecular subsystem exhibit the dimerization gap of approximately 0.5~eV, confirming the estimates from optical measurements \cite{Dressel04,Faltermeier07,Ferber14}.  Hence, the overall cation-derived character of \etcn\ band structure around the Fermi level is evident. Even though the band structure shape of the whole system around the Fermi level is fully determined by the molecular subsystem (as confirmed by the rigid band shift) - the position of the Fermi level alone is determined by the charge transfer (doping) between the anion and cation subsystems - showing strong coupling between the two. The importance of their interaction (through molecular band doping - i.e. charge transfer) is manifested through the system properties such as energy differences of the various structural configurations and phonon frequencies.

\begin{figure}
\renewcommand{\thefigure}{S\arabic{figure}}
\centering
\includegraphics[clip=true,width=0.8\columnwidth]{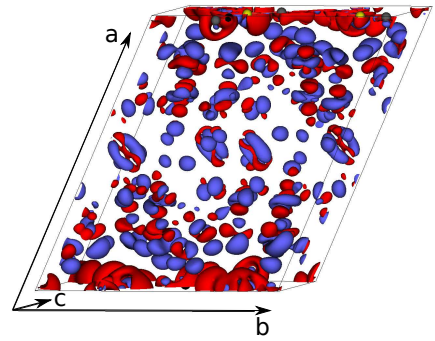}
\caption{DFT cation-anion interaction-induced charge distribution of \etcn: the BEDT-TTF molecules (regions marked in blue) donate electrons via hydrogen bonds to the acceptor anion layers (regions marked in red).
}
\label{fig:charge distribution whole}
\end{figure}

\begin{figure}
\renewcommand{\thefigure}{S\arabic{figure}}
\centering
\includegraphics[clip=true,width=0.8\columnwidth]{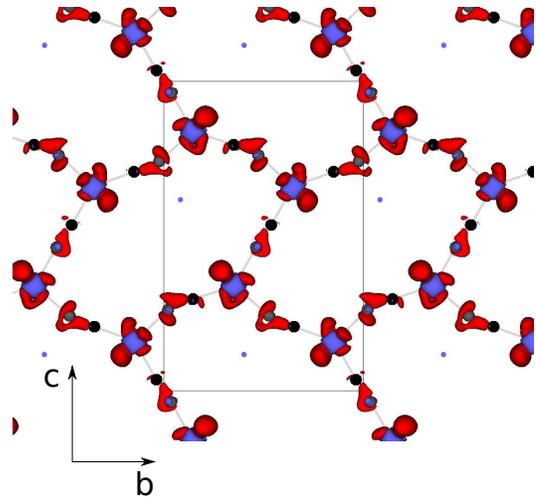}\caption{DFT cation-anion interaction-induced charge distribution of the Cu$_2$(CN)$_3^-$ anion layer in the
$bc$ plane. The electron density is enhanced (regions marked in red) at cyanide sites, much more along the chain $b$-axis than along the bridging $c$-axis. The largest electron density is concentrated in the vicinity of Cu sites; its distribution differs at Cu sites with two distinct triangular coordination which interchange along the chains in the $b$ direction: Cu atom coordinated with two N atoms and one C atom, or Cu atom coordinated with two C atoms and one N atom.
}
\label{fig:charge distribution anion}
\end{figure}

\newpage
\renewcommand{\baselinestretch}{1.0}

%\section{Electronic structure}
\newpage

\begin{table*}
\renewcommand{\thetable}{S \Roman{table}}
\caption{Vibrational modes of \etcn\ in the far- and mid-infrared spectral range up to frequencies of approximately 100~THz obtained by {\it ab-initio} phonon calculations based on density-functional-theory (DFT). C1B0 indicates the lowest-energy configuration with parallel cyanides as shown in Fig.~4 of the main paper, while C2B0 refers to the configuration with reversed $c_2$ cyanide which is higher in energy by 15~meV.\\
%Only selected vibrational modes are described.\\
\label{tab:1}}
\begin{tabular}{r | rr | rr || l}
 & \multicolumn{2}{c}{C1B0 } & \multicolumn{2}{c||}{C2B0} & \\
number & \multicolumn{4}{c||}{Frequency } &  comments\\
 & (THz) & (cm$^{-1}$)  & (THz) & (cm$^{-1}$) & \\
\hline\hline
1   & (-0.004) & 0 & (-0.012) & 0 & error of calculation yields negative energies close to zero\\
2   & (-0.003) & 0 & (-0.003) & 0 & \\
3   & (-0.002) & 0 & (-0.001) & 0 & \\
\hline
4   & 0.729  & 24.32 & 0.731 & 24.38 & {$b$-axis} shear and rocking motion of anion layer; \\
  &  &  &  &  & BEDT-TTF motion in the $ac$ plane in phase within dimer; \\
  &  &  &  &  & gliding of dimers along their extended axes \\
	&  &  &  &  & with respect to each other\\
\hline
5   & 0.998  & 33.28 & 0.995 & 33.19 & \underline{$c$-axis} shear movement of anion layer with respect to BEDT-TTF; \\
  	&  &  &  &  & breathing BEDT-TTF motion in counterphase to each other within dimer;\\
		&  &  &  &  & twisting in-phase motion of BEDT-TTF dimers \\
6   & 1.022  & 34.09 & 1.026 & 34.22 & out-of-plane vibration of anions; tilting of dimers\\
& & & & & in-phase oscillation of BEDT-TTF\\
7   & 1.054  & 35.16 & 1.062 & 35.42 & out-of-plane vibration of anions; wagging of ethylene groups\\
8   & 1.134  & 37.81 & 1.143 & 38.14 & waving of anion layer along \underline{$c$-axis};%Cu-CN-Cu out-of-phase;
\\
		&  &  &  &  & breathing BEDT-TTF motion in counterphase to each other within dimer; \\
    &  &  &  &	& twisting in-phase motion of BEDT-TTF dimers; flipping of ethylene groups\\
9   & 1.146  & 38.23 & 1.158 & 38.63 & \underline{$b$-axis} shear movement of anion layer with respect to BEDT-TTF,\\
  &  &  &  &  & bending in-phase motion of BEDT-TTF dimers\\
10  & 1.155  & 38.51 & 1.170 & 39.02 & weak \underline{$a$-axis} motion of anions; wagging of ethylene endgroups \\
11  & 1.215  & 40.52 & 1.205 & 40.18 & \underline{$b$-axis} shear movement of anion layer with respect to BEDT-TTF,\\
  &  &  &  &  & bending in-phase motion of BEDT-TTF dimers\\
12  & 1.229  & 41.00 &  &  & mainly \underline{$a$-axis} motion of anions \\
  &  &  &  &  & gliding of BEDT-TTF within dimer \\
	&  &  &  1.238 & 41.30 & \underline{$b$-axis} shear movement of anion layer with respect to BEDT-TTF,\\
	&  &  &  &  & bending in-phase motion of BEDT-TTF dimers\\
13  & 1.267  & 42.26 & 1.273 & 42.45 & waving of anion layer;\\
  &  &  &  &  & tilting of BEDT-TTF\\
\hline
14  & 1.380  & 46.03 & 1.380 & 46.03 & sheer motion of coupled BEDDT-TTF and anions\\
15  & 1.440  & 48.04 & 1.438 & 47.97 & strong out-of-plane motion of anions coupled to BEDT-TTF\\
16  & 1.463  & 48.81 & 1.466 & 48.90 & \underline{$b$-axis} motion of anion layer, tilting of bridging CN c$_1$; \\
  &  &  &  &  & rocking of BEDT-TTF dimers \\
\hline
17  & 1.514  & 50.51 & 1.525 & 50.87 & wagging of BEDT-TTF, $c$-axis motion of anions\\
18  & 1.549  & 51.68 & 1.554 & 51.83 & wagging of BEDT-TTF, $b$-axis motion of anions\\
19  & 1.606  & 53.58 & 1.596 & 53.24 & gliding of BEDT-TTF, $a$-axis motion of anions\\
20  & 1.629  & 54.32 & 1.634 & 54.51 & tilting and wagging of BEDT-TTF, $b$-axis motion of anions\\
21  & 1.641  & 54.74 & 1.645 & 54.88 & torsion motion of BEDT-TTF, $b$-axis motion of anions\\
22  & 1.683  & 56.13 & 1.689 & 56.32 & vibration of BEDT-TTF, $c$-axis motion of anions\\
\hline
23  & 1.746  & 58.24 & 1.753 & 58.46 & coupled out-of-plane anion motion and BEDT-TTF vibration\\
24  & 1.753  & 58.48 & 1.755 & 58.54 & \\
25  & 1.786  & 59.59 & 1.781 & 59.41 & \\
26  & 1.853  & 61.80 & 1.844 & 61.51 & mainly $c$-axis vibration of anions coupled to BEDT-TTF vibration\\
\hline
27  & 1.941  & 64.76 & 1.955 & 65.20 & 27-40: Coupled anion motion with BEDT-TTF vibration;\\
    &  &  &  &  &  wagging of ethylene endgroups; wagging of Cu bonds, bending of cyanide links\\
28  & 2.000  & 66.70 & 1.997 & 66.61 & \\
29  & 2.040  & 68.06 & 2.039 & 68.00 & \\
30  & 2.090  & 69.72 & 2.092 & 69.77 & \\
31  & 2.122  & 70.77 & 2.127 & 70.96 & \\
32  & 2.157  & 71.96 & 2.167 & 72.29 & \\
33  & 2.164  & 72.18 & 2.169 & 73.34 & \\
34  & 2.280  & 76.05 & 2.288 & 76.32 & \\
\end{tabular}
\end{table*}
\begin{table*}
\addtocounter {table} {-1}
\renewcommand{\thetable}{S \Roman{table}}
\caption{Continued}
\begin{tabular}{r | rr | rr || l}
 & \multicolumn{2}{c}{C1B0 } & \multicolumn{2}{c||}{C2B0} & \\
number & \multicolumn{4}{c||}{Frequency } &  comments\\
 & (THz) & (cm$^{-1}$)  & (THz) & (cm$^{-1}$) & \\
\hline\hline
%\hline
35  & 2.290  & 76.39 & 2.295 & 76.56 & \\
36  & 2.333  & 77.80 & 2.338 & 77.99 & \\
37  & 2.435  & 81.21 & 2.442 & 81.45 & \\
38  & 2.474  & 82.51 & 2.483 & 82.82 & \\
39  & 2.484  & 82.85 & 2.493 & 83.15 & \\
40  & 2.539  & 84.69 & 2.545 & 84.89 & \\
41  & 2.555  & 85.21 & 2.561 & 85.43 & 41-62: rotation of Cu bonds and twisting of cyano bonds; rocking of ethylene groups \\
42  & 2.566  & 85.59 & 2.573 & 85.81 & \\
43  & 2.594  & 86.52 & 2.600 & 86.72 & \\
44  & 2.630  & 87.71 & 2.641 & 88.10 & \\
45  & 2.727  & 90.96 & 2.727 & 90.97 & \\
46  & 2.748  & 91.68 & 2.752 & 91.79 & \\
%\hline
47  & 2.853  & 95.18 & 2.857 & 95.28 & \\
48  & 2.883  & 96.18 & 2.892 & 96.47 & \\
49  & 2.969  & 99.04 & 2.965 & 98.89 & \\
50  & 2.976  & 99.25 & 2.996 & 99.95 & \\
51  & 3.030  & 101.06 & 3.039 & 101.37 & \\
52  & 3.044  & 101.54 & 3.052 & 101.81 & \\
53  & 3.055  & 101.91 & 3.071 & 102.43 & \\
54  & 3.119  & 104.02 & 3.125 & 104.25 & \\
55  & 3.271  & 109.10 & 3.281 & 109.46 & \\
56  & 3.306  & 110.27 & 3.311 & 110.43 & \\
57  & 3.446  & 114.95 & 3.470 & 115.74 & \\
58  & 3.490  & 116.40 & 3.486 & 116.29 & \\
59  & 3.653  & 121.85 & 3.649 & 121.70 & \\
60  & 3.714  & 123.89 & 3.716 & 123.97 & \\
61  & 3.720  & 124.78 & 3.741 & 124.80 & \\
62  & 3.757  & 125.32 & 3.744 & 124.88 & \\
\hline
63  & 3.858  & 128.70 & 3.873 & 129.78 & 63-68: bending and stretching of anions coupled to BEDT-TTF\\
64  & 3.870  & 129.08 & 3.873 & 129.20 & \\
65  & 3.967  & 132.34 & 3.977 & 132.66 & \\
66  & 3.981  & 132.80 & 3.985 & 132.94 & \\
67  & 3.992  & 133.15 & 4.001 & 133.47 & \\
68  & 4.009  & 133.72 & 4.020 & 134.10 & \\
\hline
69  & 4.050  & 135.09 & 4.054 & 135.22 & twist of cyanides, rotation motion of Cu coupled to BEDT-TTF \\
70  & 4.154  & 138.57 & 4.182 & 139.49 & out-of-plane bending of cyanides, mainly c$_1$ and c$_2$, coupled to BEDT-TTF\\
71  & 4.280  & 142.77 & 4.283 & 142.88 & out-of-plane bending of cyanides, mainly b$_3$ and b$_4$; coupled to BEDT-TTF\\
72  & 4.306  & 143.63 & 4.350 & 145.09 & out-of-plane bending of cyanides coupled to BEDT-TTF\\
\hline
73  & 4.755  & 158.60 & 4.702 & 156.84 & 73-78: anion motion coupled to BEDT-TTF;\\
    &  &  &  &  & bending and stretching of cyanides; mainly $c$-direction\\
74  & 4.836  & 161.32 & 4.760 & 158.76 & \\
75  & 4.848  & 161.70 & 4.850 & 161.76 & \\
76  & 4.950  & 165.11 & 4.945 & 164.93 & \\
77  & 4.983  & 166.22 & 4.976 & 165.99 & \\
78  & 5.065  & 168.96 & 5.015 & 167.28 & 78-83: anion motion coupled to BEDT-TTF; bending and stretching of cyanides\\
79  & 5.159  & 172.09 & 5.227 & 174.34 & \\
80  & 5.407  & 180.36 & 5.411 & 180.50 & 80-83: anion motion coupled to twisting BEDT-TTF; \\
& & & & & bending and stretching of cyanides\\
81  & 5.414  & 180.60 & 5.414 & 180.59 & \\
82  & 5.491  & 183.20 & 5.495 & 183.30 & \\
83  & 5.504  & 183.59 & 5.505 & 183.63 & \\
\end{tabular}
\end{table*}
\begin{table*}
\addtocounter {table} {-1}
\renewcommand{\thetable}{S \Roman{table}}
\caption{Continued}
\begin{tabular}{r | r r | rr || l}
& \multicolumn{2}{c}{C1B0 } & \multicolumn{2}{c||}{C2B0} & \\
number & \multicolumn{4}{c||}{Frequency } & comment\\
 & (THz) & (cm$^{-1}$)  & (THz) & (cm$^{-1}$) & \\
\hline\hline
84  & 5.844  & 194.94 & 5.894 & 196.61 & 84-88: twisting and stretching anion layer coupled to ethylene endgroups\\
85  & 5.963  & 198.91 & 6.016 & 200.68 & \\
86  & 6.177  & 206.05 & 6.020 & 200.80 & \\
87  & 6.455  & 215.32 & 6.536 & 218.00 & \\
88  & 6.741  & 224.86 & 6.774 & 225.96 & \\
\hline
89  & 6.916  & 230.69 & 6.921 & 230.86 & 89-96: bending of BEDT-TTF; no anion motion\\
90  & 6.946  & 231.69 & 6.951 & 231.87 & \\
91  & 7.257  & 242.06 & 7.263 & 242.28 & \\
92  & 7.263  & 242.26 & 7.276 & 242.69 & \\
93  & 7.302  & 243.59 & 7.310 & 243.82 & \\
94  & 7.313  & 243.92 & 7.325 & 244.34 & \\
95  & 7.422  & 247.56 & 7.436 & 248.03 & \\
96  & 7.440  & 248.16 & 7.455 & 248.66 & \\
\hline
97  & 7.636  & 254.72 & 7.641 & 254.86 & 97-104: wagging movement of BEDT-TTF coupled to-out-plane motion of anions\\
98  & 7.661  & 255.54 & 7.644 & 254.97 & \\
99  & 7.699  & 256.81 & 7.701 & 256.87 & \\
100 & 7.730  & 257.84 & 7.729 & 257.80 & \\
101 & 7.774  & 259.30 & 7.781 & 259.56 & \\
102 & 7.791  & 259.87 & 7.798 & 260.12 & \\
103 & 7.867  & 262.40 & 7.843 & 261.62 & \\
104 & 7.883  & 262.95 & 7.885 & 263.02 & \\
\hline
105 & 7.984  & 266.31 & 7.947 & 265.07 & 105-106: out of plane twist of cyanides, coupled to BEDT-TTF vibration\\
    &  &  &  &  & mainly c$_1$ and b$_2$\\
106 & 8.029  & 266.31 & 7.968 & 265.77 & mainly c$_2$ and b$_4$\\
\hline
107 & 8.089  & 269.81 & 8.093 & 269.96 & 107-114: rocking of ethylene endgroups coupled to anions\\
108 & 8.103  & 270.29 & 8.112 & 270.60 & \\
109 & 8.157  & 272.09 & 8.128 & 271.12 & \\
110 & 8.210  & 273.84 & 8.201 & 273.55 & \\
111 & 8.392  & 279.73 & 8.408 & 280.46 & \\
112 & 8.400  & 280.21 & 8.416 & 280.74 & \\
113 & 8.443  & 281.63 & 8.458 & 282.13 & \\
114 & 8.485  & 283.04 & 8.494 & 283.32 & \\
\hline
115 & 8.769  & 292.49 & 8.703 & 290.31 & 115-120: twisting of cyanides coupled to rocking ethylene groups\\
116 & 8.799  & 293.51 & 8.783 & 292.96 & \\
117 & 8.836  & 294.74 & 8.835 & 294.72 & \\
118 & 8.897  & 296.76 & 8.895 & 296.79 & \\
119 & 8.939  & 298.17 & 8.937 & 298.11 & \\
120 & 9.026  & 301.07 & 8.956 & 298.73 & \\
\hline
\end{tabular}
\end{table*}
\begin{table*}
\addtocounter {table} {-1}
\renewcommand{\thetable}{S \Roman{table}}
\caption{Continued}
\begin{tabular}{r | r r | rr || l}
& \multicolumn{2}{c}{C1B0 } & \multicolumn{2}{c||}{C2B0} & \\
number & \multicolumn{4}{c||}{Frequency } & comment\\
 & (THz) & (cm$^{-1}$)  & (THz) & (cm$^{-1}$) & \\
\hline\hline
121 & 9.280  & 309.56 & 9.282 & 309.61 & 121-125: various BEDT-TTF vibrations coupled with anion vibrations\\
122 & 9.286  & 309.75 & 9.286 & 309.75 & \\
123 & 9.309  & 310.52 & 9.314 & 310.69 & 123-129: twisting of cyanides\\
124 & 9.361  & 312.26 & 9.328 & 311.16 & \\
125 & 9.390  & 313.23 & 9.389 & 313.17 & 125-129: twisting of cyanides coupled with weak rocking of ethylene endgroups\\
126 & 9.496  & 316.74 & 9.589 & 319.27 & mainly b$_3$ and b$_1$, and c$_1$ and c$_2$\\
127 & 9.635  & 321.38 & 9.746 & 325.09 & \\
128 & 9.883  & 329.66 & 9.823 & 327.66 & mainly b$_4$ and b$_2$, and c$_1$ and c$_2$\\
%\hline
129 & 10.190 & 339.89 & 10.037 & 334.80 & additional BEDT-TTF tilting\\
\hline
130 & 10.293 & 343.35 & 10.184 & 339.69 & 130-144: twisting and shuffling of BEDT-TTF coupled with anion vibrations;\\
    &        &        &        &    & rocking of ethylene endgroups
%in phase
\\
131 & 10.320 & 344.25 & 10.288 & 343.16 & \\
132 & 10.338 & 344.84 & 10.316 & 343.16 & %twisting out of phase
\\
133 & 10.356 & 345.45 & 10.338 & 344.83 & \\
134 & 10.382 & 346.30 & 10.356 & 345.43 & 134-144: twisting of cyanides\\
135 & 10.418 & 347.49 & 10.381 & 346.26 & \\
136 & 10.420 & 347.57 & 10.392 & 346.64 & \\
137 & 10.427 & 347.80 & 10.427 & 347.80 & \\
138 & 10.485 & 349.73 & 10.428 & 347.85 & \\
139 & 10.520 & 350.90 & 10.511 & 350.60 & \\
140 & 10.562 & 352.30 & 10.550 & 351.92 & \\
141 & 10.583 & 353.03 & 10.566 & 352.44 & \\
142 & 10.602 & 353.66 & 10.595 & 353.42 & \\
%\hline
143 & 10.703 & 357.03 & 10.673 & 356.00 & 143-144: strong twisting of cyanides\\
144 & 11.211 & 373.95 & 11.301 & 376.96 & \\
\hline
145 & 11.684 & 389.72 & 11.674 & 389.42 & 145-148: antisymmetric breathing modes of BEDT-TTF rings
%, out of phase within dimer
\\
146 & 11.696 & 390.13 & 11.687 & 389.83 & \\
147 & 11.729 & 391.25 & 11.720 & 390.94 & \\
148 & 11.743 & 391.72 & 11.734 & 391.42 & \\
\hline
149 & 12.982 & 433.04 & 12.969 & 432.60 & 149-160: symmetric breathing modes of BEDT-TTF rings\\
150 & 13.016 & 434.18 & 13.005 & 433.79 & %breathing mode of rings in one BEDT-TTF molecule
\\
151 & 13.021 & 434.32 & 13.007 & 433.88 & %breathing modes of rings, all four BEDT-TTF in phase
\\
152 & 13.046 & 435.16 & 13.034 & 434.75 & \\
153 & 13.572 & 452.73 & 13.564 & 452.45 & \\
154 & 13.582 & 453.05 & 13.574 & 452.79 & \\
155 & 13.619 & 454.29 & 13.614 & 454.11 & \\
156 & 13.624 & 454.46 & 13.616 & 454.18 & \\
157 & 13.661 & 455.69 & 13.651 & 455.36 & \\
158 & 13.676 & 456.17 & 13.669 & 455.96 & \\
159 & 13.746 & 458.52 & 13.737 & 458.23 & \\
160 & 13.749 & 458.61 & 13.739 & 458.27 & \\
\end{tabular}
\end{table*}
\begin{table*}
\addtocounter {table} {-1}
\renewcommand{\thetable}{S \Roman{table}}
\caption{Continued}
\begin{tabular}{r | r r | rr || l}
& \multicolumn{2}{c}{C1B0 } & \multicolumn{2}{c||}{C2B0} & \\
number & \multicolumn{4}{c||}{Frequency } & comment\\
 & (THz) & (cm$^{-1}$)  & (THz) & (cm$^{-1}$) & \\
\hline\hline
161 & 13.762 & 459.04 & 13.758 & 458.91 & 161-170: symmetric breathing modes of BEDT-TTF rings\\
162 & 13.780 & 459.64 & 13.767 & 459.23 & \\
163 & 13.800 & 460.30 & 13.773 & 459.43 & \\
164 & 13.828 & 461.24 & 13.803 & 460.43 & \\
165 & 13.844 & 461.78 & 13.836 & 461.51 & \\
166 & 13.921 & 464.36 & 13.916 & 464.20 & \\
167 & 13.978 & 466.27 & 13.974 & 466.11 & \\
168 & 13.997 & 466.89 & 13.991 & 466.68 & breathing of inner rings; dimers in phase \\
169 & 14.085 & 469.82 & 14.077 & 469.57 & breathing of inner rings in all four BEDT-TTF\\
170 & 14.215 & 474.46 & 14.140 & 471.64 & vibration of cyanide group as a whole, mainly b$_4$ and b$_2$ \\
    &        &        &        &        & coupled with weak rocking of ethylene endgroups\\
\hline
171 & 14.278 & 476.27 & 14.269 & 475.98 & 171-172: antisymmetric breathing of inner rings within dimer; \\
    &        &        &        &        &   rocking of central C=C \\
172 & 14.284 & 476.45 & 14.275 & 476.16 & \\
\hline
173 & 14.346 & 478.52 & 14.338 & 478.25 & 173-174: symmetric breathing of inner rings within dimer; \\
    &  &  &  &  &   rocking of central C=C\\
174 & 14.478 & 482.94 & 14.416 & 480.87 & coupled with anion vibrations\\
\hline
175 & 14.506 & 483.88 & 14.472 & 482.74 & 175-180: Cu-CN-Cu motion; rocking of outer C=C bonds of BEDT-TTF;\\
&  &  &  &  &   dominant motion of b$_3$ and also b$_4$; dominant rocking of one outer C=C \\
176 & 14.590 & 485.66 & 14.552 & 485.39 & dominant motion of b$_1$ and also b$_2$\\
%\hline
177 & 14.577 & 486.23 & 14.560 & 485.67 & \\
178 & 14.589 & 486.64 & 14.571 & 486.05 & \\
%\hline
179 & 14.595 & 486.83 & 14.575 & 486.17 & dominant motion of b$_1$ and b$_4$\\
180 & 14.604 & 487.15 & 14.649 & 488.63 & dominant motion of b$_1$ and b$_2$ \\
\hline
181 & 14.667 & 489.24 & 14.650 & 488.68 & 181-184: tilting motion of one outer C=C bond and outer ring of BEDT-TTF\\
182 & 14.670 & 489.34 & 14.656 & 488.86 & \\
183 & 14.672 & 489.40 & 14.657 & 488.90 & \\
184 & 14.675 & 489.45 & 14.807 & 493.92 & \\
\hline
185 & 15.215 & 507.51 & 15.243 & 508.46 & 185-186: vibration of cyanides with respect copper ions: Cu-CN-Cu;\\
& & & & & c$_1$ and c$_2$ are in phase \\
186 & 15.402 & 513.74 & 15.447 & 515.26 & c$_1$ and c$_2$ are out of phase\\
\hline
187 & 17.997 & 600.32 & 17.997 & 600.31 & 187-200: rocking of ethylene endgroups \\
188 & 18.000 & 600.42 & 18.000 & 600.41 & \\
189 & 18.010 & 600.75 & 18.014 & 600.88 & \\
190 & 18.015 & 600.91 & 18.017 & 600.99 & \\
191 & 18.206 & 607.29 & 18.201 & 607.11 & \\
192 & 18.208 & 607.35 & 18.203 & 607.17 & \\
193 & 18.230 & 608.09 & 18.212 & 607.50 & \\
194 & 18.235 & 608.27 & 18.214 & 607.54 & \\
%\hline
195 & 19.128 & 638.05 & 19.128 & 638.03 & \\
196 & 19.133 & 638.19 & 19.133 & 638.22 & \\
197 & 19.148 & 638.71 & 19.149 & 638.75 & \\
198 & 19.149 & 638.74 & 19.153 & 638.87 & \\
199 & 19.222 & 641.16 & 19.218 & 641.03 & \\
200 & 19.227 & 641.35 & 19.221 & 641.15 & \\
\end{tabular}
\end{table*}
\begin{table*}
\addtocounter {table} {-1}
\renewcommand{\thetable}{S \Roman{table}}
\caption{Continued}
\begin{tabular}{r | r r | rr || l}
& \multicolumn{2}{c}{C1B0 } & \multicolumn{2}{c||}{C2B0} & \\
number & \multicolumn{4}{c||}{Frequency } & comment\\
 & (THz) & (cm$^{-1}$)  & (THz) & (cm$^{-1}$) & \\
\hline\hline
201 & 19.231 & 641.49 & 19.223 & 641.19 & 201-202: rocking of ethylene endgroups\\
202 & 19.242 & 641.84 & 19.225 & 641.29 & \\
\hline
203 & 22.346 & 745.38 & 22.327 & 744.76 & 203-210: rocking of C-C and ethylene endgroups; \\
& & & & & twist of outer ring of BEDT-TTF\\
204 & 22.369 & 746.14 & 22.351 & 745.55 & \\
205 & 22.385 & 746.69 & 22.366 & 746.06 & \\
206 & 22.396 & 747.04 & 22.377 & 746.43 & \\
%\hline
207 & 22.517 & 751.09 & 22.491 & 750.21 & twist of outer ring at one BEDT-TTF side \\
208 & 22.523 & 751.30 & 22.504 & 750.65 & \\
209 & 22.529 & 751.50 & 22.506 & 750.73 & \\
210 & 22.556 & 752.40 & 22.535 & 751.68 & \\
\hline
211 & 22.720 & 757.87 & 22.718 & 757.79 & 211-214: longitudinal translation of central C=C bond of BEDT-TTF; \\
& & & & & rocking of ethylene endgroups\\
212 & 22.730 & 758.18 & 22.728 & 758.13 & \\
213 & 22.743 & 758.64 & 22.739 & 758.48 & \\
214 & 22.749 & 758.84 & 22.745 & 758.68 & \\
\hline
215 & 24.939 & 831.87 & 24.940 & 831.90 & 214-218: transverse translation of central C=C bond of BEDT-TT; \\
& & & & & rocking of ethylene endgroups\\
216 & 24.945 & 832.08 & 24.947 & 832.13 & \\
217 & 24.970 & 832.91 & 24.969 & 832.87 & \\
218 & 24.984 & 833.35 & 24.983 & 833.34 & \\
\hline
219 & 25.112 & 837.63 & 25.072 & 836.32 & 219-222: symmetric translations of outer C=C and breathing of rings;\\
& & & & & rocking of ethylene endgroups\\
220 & 25.141 & 838.60 & 25.102 & 837.32 & \\
221 & 25.149 & 838.89 & 25.108 & 837.52 & \\
222 & 25.191 & 840.28 & 25.150 & 838.90 & \\
\hline
223 & 25.327 & 844.81 & 25.289 & 843.54 & 223-226: asymmetric translations of outer C=C and breathing of inner rings;\\
& & & & & rocking of ethylene endgroups\\
224 & 25.379 & 846.57 & 25.343 & 845.34 & \\
225 & 25.392 & 846.97 & 25.354 & 845.71 & \\
226 & 25.406 & 847.46 & 25.369 & 846.22 & \\
\hline
227 & 25.931 & 864.96 & 25.929 & 864.90 & 227-240 rocking of ethylene endgroups\\
%wagging of CH$_2$\\
228 & 25.954 & 865.74 & 25.949 & 865.57 & \\
229 & 25.959 & 865.91 & 25.960 & 865.94 & \\
230 & 25.978 & 866.53 & 25.973 & 866.39 & \\
%\hline
231 & 26.182 & 873.34 & 26.171 & 872.98 & \\
232 & 26.190 & 873.62 & 26.174 & 873.07 & \\
233 & 26.202 & 873.99 & 26.196 & 873.81 & \\
234 & 26.226 & 874.82 & 26.218 & 874.54 & \\
%\hline
235 & 27.086 & 903.51 & 27.076 & 903.16 & \\
236 & 27.096 & 903.84 & 27.085 & 903.46 & \\
237 & 27.108 & 903.84 & 27.103 & 904.07 & \\
238 & 27.109 & 904.26 & 27.104 & 904.08 & \\
239 & 27.199 & 907.27 & 27.174 & 906.42 & \\
240 & 27.214 & 907.76 & 27.188 & 906.91 & \\
\end{tabular}
\end{table*}
\begin{table*}
\addtocounter {table} {-1}
\renewcommand{\thetable}{S \Roman{table}}
\caption{Continued}
\begin{tabular}{r | r r | rr || l}
& \multicolumn{2}{c}{C1B0 } & \multicolumn{2}{c||}{C2B0} & \\
number & \multicolumn{4}{c||}{Frequency } & comment\\
 & (THz) & (cm$^{-1}$)  & (THz) & (cm$^{-1}$) & \\
\hline\hline
241 & 27.237 & 908.52 & 27.212 & 907.69 & 241-242: rocking of ethylene endgroups\\
242 & 27.254 & 909.09 & 27.228 & 908.24 & \\
\hline
243 & 28.552 & 952.38 & 28.500 & 950.66 & 243-262: twist of C=C bonds of BEDT-TTF\\
244 & 28.566 & 952.85 & 28.514 & 951.11 & \\
245 & 28.593 & 953.75 & 28.540 & 952.01 & \\
246 & 28.605 & 954.15 & 28.551 & 952.35 & \\
247 & 28.722 & 958.07 & 28.669 & 956.30 & \\
248 & 28.742 & 958.72 & 28.689 & 956.95 & \\
249 & 28.781 & 960.02 & 28.730 & 958.32 & \\
250 & 28.801 & 960.70 & 28.750 & 958.99 & \\
%\hline
251 & 29.238 & 975.27 & 29.205 & 974.19 & \\
%C-C streching vibrations\\
252 & 29.240 & 975.35 & 29.212 & 974.39 & \\
253 & 29.256 & 975.89 & 29.220 & 974.68 & \\
254 & 29.261 & 976.04 & 29.229 & 974.98 & \\
%\hline
255 & 29.363 & 979.46 & 29.345 & 978.83 & strong twist of central C=C bonds \\
256 & 29.383 & 980.11 & 29.360 & 979.36 & \\
257 & 29.398 & 980.61 & 29.378 & 979.93 & \\
258 & 29.412 & 981.09 & 29.381 & 980.05 & \\
%\hline
259 & 29.532 & 985.08 & 29.455 & 982.50 & \\
%C-C streching vibrations,%CH$_2$ tilting modes\\
260 & 29.538 & 985.27 & 29.454 & 982.51 & \\
261 & 29.547 & 985.58 & 29.465 & 982.84 & \\
262 & 29.554 & 985.83 & 29.478 & 983.27 & \\
\hline
263 & 33.235 & 1108.60 & 33.233 & 1108.53 & 263-270: ethylene endgroups tilting modes \\
264 & 33.239 & 1108.74 & 33.239 & 1108.74 & \\
265 & 33.249 & 1109.05 & 33.243 & 1108.87 & \\
266 & 33.251 & 1109.12 & 33.246 & 1108.97 & \\
267 & 33.408 & 1114.39 & 33.387 & 1113.67 & \\
268 & 33.420 & 1114.76 & 33.395 & 1113.93 & \\
269 & 33.427 & 1117.99 & 33.404 & 1114.23 & \\
270 & 33.438 & 1115.39 & 33.413 & 1114.54 & \\
\hline
271 & 34.878 & 1163.40 & 34.881 & 1163.51 & 271-280: ethylene endgroups wagging modes \\
272 & 34.892 & 1163.88 & 34.895 & 1163.96 & \\
273 & 34.913 & 1164.60 & 34.906 & 1164.35 & \\
274 & 34.930 & 1165.15 & 34.919 & 1164.78 & \\
275 & 35.051 & 1169.17 & 34.997 & 1167.39 & \\
276 & 35.060 & 1169.48 & 35.013 & 1167.91 & \\
277 & 35.068 & 1169.75 & 35.021 & 1168.18 & \\
278 & 35.090 & 1170.49 & 35.048 & 1169.07 & \\
%\hline
279 & 37.592 & 1253.93 & 37.586 & 1253.72 & \\
280 & 37.601 & 1254.23 & 37.602 & 1254.28 & \\
\end{tabular}
\end{table*}
\begin{table*}
\addtocounter {table} {-1}
\renewcommand{\thetable}{S \Roman{table}}
\caption{Continued}
\begin{tabular}{r | r r | rr || l}
& \multicolumn{2}{c}{C1B0 } & \multicolumn{2}{c||}{C2B0} & \\
number & \multicolumn{4}{c||}{Frequency } & comment\\
 & (THz) & (cm$^{-1}$)  & (THz) & (cm$^{-1}$) & \\
\hline\hline
281 & 37.615 & 1254.70 & 37.605 & 1254.37 & 281-294: ethylene endgroups wagging modes \\
282 & 37.621 & 1254.59 & 37.610 & 1254.53 & \\
283 & 37.669 & 1256.49 & 37.634 & 1255.35 & \\
284 & 37.681 & 1256.89 & 37.637 & 1255.44 & \\
285 & 37.706 & 1257.23 & 37.679 & 1256.85 & \\
286 & 37.722 & 1258.26 & 37.685 & 1257.06 & \\
%\hline
287 & 38.153 & 1272.66 & 38.151 & 1272.59 & \\
288 & 38.170 & 1273.21 & 38.166 & 1273.08 & \\
289 & 38.178 & 1273.47 & 38.177 & 1273.45 & \\
290 & 38.198 & 1274.14 & 38.195 & 1274.06 & \\
291 & 38.611 & 1287.92 & 38.614 & 1288.04 & \\
292 & 38.633 & 1288.67 & 38.617 & 1288.14 & \\
293 & 38.639 & 1288.86 & 38.640 & 1288.90 & \\
294 & 38.665 & 1289.71 & 38.651 & 1289.27 & \\
\hline
295 & 41.479 & 1383.58 & 41.583 & 1387.06 & 295-297: symmetric C=C vibrations; $\nu_3$(a$_{\rm g}$)  in D$_{\rm 2h}$ molecular symmmetry;\\
& & & & & two dimers are out of phase \\
296 & 41.489 & 1383.92 & 41.594 & 1387.43 & \\
%antisymmetric C=C vibrations; $\nu_{27}$(b$_{\rm 1u}$) in D$_{\rm 2h}$ molecular symmmetry\\
%& & & & & two dimers are out of phase \\
297 & 41.673 & 1390.05 & 41.774 & 1393.45 & \\
%antisymmetric C=C vibrations; $\nu_{27}$(b$_{\rm 1u}$)  in D$_{\rm 2h}$ molecular symmmetry\\
%& & & & & two dimers are out of phase, inverted phase relation between dimers\\
\hline
298 & 42.346 & 1412.50 & 42.340 & 1412.30 & 298-318: ethylene endgroups scissoring modes \\
299 & 42.383 & 1413.75 & 42.370 & 1413.30 & \\
300 & 42.383 & 1413.76 & 42.379 & 1413.62 & \\
301 & 42.393 & 1414.08 & 42.380 & 1413.66 & \\
302 & 42.535 & 1418.81 & 42.509 & 1417.93 & \\
303 & 42.560 & 1419.64 & 42.533 & 1418.74 & \\
304 & 42.591 & 1420.68 & 42.573 & 1420.07 & \\
305 & 42.594 & 1420.79 & 42.583 & 1420.40 & \\
306 & 42.604 & 1421.11 & 42.590 & 1420.45 & \\
307 & 42.614 & 1421.43 & 42.610 & 1421.32 & \\
308 & 42.621 & 1421.69 & 42.620 & 1421.63 & \\
309 & 42.633 & 1422.07 & 42.627 & 1421.88 & \\
\hline
310 & 42.748 & 1425.91 & 42.847 & 1429.22 & antisymmetric C=C vibrations; $\nu_{27}$(b$_{\rm 1u}$) in D$_{\rm 2h}$ molecular symmmetry;\\
& & & & & out of phase within dimer; two dimers are in-phase\\
311 & 42.826 & 1428.54 & 42.928 & 1431.92 & antisymmetric C=C vibrations; $\nu_{27}$(b$_{\rm 1u}$) in D$_{\rm 2h}$ molecular symmmetry;\\
& & & & & out of phase within dimer; two dimers are out-of-phase\\
%\hline
312 & 42.898 & 1430.91 & 42.960 & 1433.00 & antisymmetric C=C vibrations; $\nu_{27}$(b$_{\rm 1u}$) in D$_{\rm 2h}$ molecular symmmetry;\\
& & & & & in-phase within dimer; two dimers are out-of-phase\\
313 & 42.927 & 1431.89 & 42.974 & 1433.45 & antisymmetric C=C vibrations; $\nu_{27}$(b$_{\rm 1u}$) in D$_{\rm 2h}$ molecular symmmetry;\\
& & & & & all four BEDT-TTF molecules area in phase\\
\hline
314 & 43.036 & 1435.54 & 43.047 & 1435.88 & \\
315 & 43.063 & 1436.43 & 43.067 & 1436.57 & \\
316 & 43.085 & 1437.15 & 43.083 & 1437.10 & \\
317 & 43.089 & 1437.30 & 43.113 & 1438.10 & \\
318 & 43.120 & 1438.32 & 43.181 & 1440.35 & \\
\hline
319 & 43.989 & 1467.30 & 44.083 & 1470.45 & 319-320: symmetric C=C vibrations: $\nu_2$ in D$_{\rm 2h}$ molecular symmetry; \\
& & & & & out of phase within dimers\\
320 & 44.000 & 1467.67 & 44.094 & 1470.83 & two dimers are out-of-phase\\
\end{tabular}
\end{table*}
\begin{table*}
\addtocounter {table} {-1}
\renewcommand{\thetable}{S \Roman{table}}
\caption{Continued}
\begin{tabular}{r | r r | rr || l}
& \multicolumn{2}{c}{C1B0 } & \multicolumn{2}{c||}{C2B0} & \\
number & \multicolumn{4}{c||}{Frequency } & comment\\
 & (THz) & (cm$^{-1}$)  & (THz) & (cm$^{-1}$) & \\
\hline\hline
321 & 44.007 & 1467.92 & 44.104 & 1471.14 & symmetric C=C vibrations: $\nu_2$ in D$_{\rm 2h}$ molecular symmetry; \\
& & & & & in phase within dimers, the two dimers are out of phase\\
322 & 44.022 & 1468.43 & 44.119 & 1471.66 &  symmetric C=C vibrations: $\nu_2$ in D$_{\rm 2h}$ molecular symmetry; \\
& & & & & all four BEDT-TTF molecules area in phase\\
\hline
323 & 62.652 & 2089.86 &  &  & cyanide stretching, mainly b$_4$\\
  &  &  &  62.895 & 2097.94 & strongest motion in b$_4$ and b$_4$\\
324 & 62.812 & 2095.17 &  &  & cyanide stretching, mainly b$_1$\\
  &  &  &  62.947 & 2099.68 & strongest motion in b$_1$ and b$_2$\\
325 & 63.003 & 2101.57 &  &  & cyanide stretching, strongest motion in b$_2$\\
  &  &  &  63.012 & 2101.84 & equally strong motion in all four b-type cyanides\\
326 & 63.160 & 2106.78 &  &  & cyanide stretching, strongest motion in b$_3$\\
  &  &  &  63.131 & 2105.84 & equally strong motion in all four b-type cyanides\\
327 & 63.910 & 2131.81 & 63.940 & 2132.82 & cyanide stretching, out-of-phase vibration of c$_1$ and c$_2$\\
328 & 64.037 & 2136.06 & 64.083 & 2137.57 & cyanide stretching, in-phase vibration of c$_1$ and c$_2$\\	
\hline
329 & 89.275 & 2977.91 & 89.256 & 2977.26 & 329-344: symmetric C-H stretching vibrations in ethylene endgroups \\
330 & 89.280 & 2978.04 & 89.257 & 2977.28 & \\
331 & 89.298 & 2978.67 & 89.312 & 2979.12 & \\
332 & 89.306 & 2978.93 & 89.337 & 2979.96 & \\
333 & 89.423 & 2982.84 & 89.449 & 2983.69 & \\
334 & 89.431 & 2983.08 & 89.459 & 2984.97 & \\
335 & 89.441 & 2983.44 & 89.465 & 2984.22 & \\
336 & 89.467 & 2984.28 & 89.483 & 2984.84 & \\
337 & 89.601 & 2988.76 & 89.557 & 2987.28 & \\
338 & 89.623 & 2989.51 & 89.558 & 2987.34 & \\
339 & 89.630 & 2989.72 & 89.599 & 2988.71 & \\
340 & 89.644 & 2990.20 & 89.602 & 2988.80 & \\
%\hline
341 & 89.831 & 2996.44 & 89.999 & 3002.04 & \\
342 & 89.849 & 2997.04 & 90.000 & 3002.09 & \\
343 & 89.865 & 2997.58 & 90.015 & 3002.56 & \\
344 & 89.908 & 2999.01 & 90.018 & 3002.66 & \\
\hline
345 & 90.773 & 3027.87 & 90.821 & 3029.46 & 345-360: antisymmetric C-H stretching vibrations in ethylene endgroups \\
346 & 90.782 & 3028.17 & 90.834 & 3029.88 & \\
347 & 90.794 & 3028.55 & 90.836 & 3029.96 & \\
348 & 90.821 & 3029.45 & 90.855 & 3030.61 & \\
349 & 91.079 & 3038.07 & 91.037 & 3036.68 & \\
350 & 91.097 & 3038.68 & 91.042 & 3036.85 & \\
351 & 91.125 & 3039.59 & 91.071 & 3037.79 & \\
352 & 91.132 & 3039.83 & 91.087 & 3038.34 & \\
353 & 91.222 & 3042.84 & 91.238 & 3043.37 & \\
354 & 91.250 & 3043.78 & 91.241 & 3043.48 & \\
355 & 91.255 & 3043.94 & 91.272 & 3044.50 & \\
356 & 91.261 & 3044.14 & 91.277 & 3044.68 & \\
%\hline
357 & 91.888 & 3065.05 & 92.040 & 3070.14 & \\
358 & 91.891 & 3065.16 & 92.057 & 3070.68 & \\
359 & 91.893 & 3065.21 & 92.065 & 3070.67 & \\
360 & 91.917 & 3066.02 & 92.066 & 3070.98 & \\
\end{tabular}
\end{table*}


\begin{thebibliography}{99}

\bibitem{Kanoda11}
K. Kanoda and R. Kato, Annu.\ Rev.\ Condens.\ Matter Phys. {\bf 2},
167 (2011); B. J. Powell and R. H McKenzie, Rep.\ Prog.\ Phys.\ {\bf 74}, 056501 (2011).

\bibitem{Shimizu03}
Y. Shimizu, K. Miyagawa,  K. Kanoda, M. Maesato, and G. Saito, Phys.\ Rev.\ Lett.\ {\bf 91},  107001 (2003):
%\bibitem{Kurosaki05}
 Y. Kurosaki, Y. Shimizu,  K. Miyagawa, K. Kanoda, G. Saito,
Phys.\ Rev.\ Lett.\ {\bf 95}, 177001 (2005).

\bibitem{Yamashita08}
S. Yamashita, Y. Nakazawa, M. Oguni, Y. Oshima, H. Jojiri,
K. Miyagawa, and K. Kanoda, Nature Phys.\ {\bf 4}, 459 (2008).

\bibitem{Yamashita09}
M. Yamashita, N. Nakata, Y. Kasahara, T. Sasaki, N. Yoneyama,
N. Kobayashi, S. Fujimoto, T. Shibauchi, and Y. Matsuda, Nature
Phys.\ {\bf 5}, 44 (2009).

%\bibitem{Huse88}D. A. Huse and V. Elser, Phys. Rev. Lett. {\bf 60}, 2531 (1988).

%\bibitem{Capriotti99}L. Capriotti, A. E. Trumper, and S. Sorella, Phys.\ Rev.\ Lett.\ {\bf 82}, 3899 (1999).

\bibitem{Kaneko14}
R. Kaneko, S. Morita, and M. Imada,
%Gapless Spin-Liquid Phase in an Extended Spin 1/2 Triangular Heisenberg Model
J.\ Phys.\ Soc.\ Jpn.\ {\bf 83}, 093707 (2014).

\bibitem{Lee08}
P. A. Lee, Science {\bf 321}, 1306 (2008);
L. Balents, Nature {\bf 464}, 199 (2010).

\bibitem{Ng07}
T.-K. Ng and P. A. Lee, Phys. Rev. Lett. {\bf 99}, 156402 (2007).
\bibitem{Kezsmarki06}
I. Kezsm{\'a}rki, Y. Shimizu, G. Mihaly, Y. Tokura, K. Kanoda, and G. Saito, Phys. Rev. B {\bf 74}, 201101  (2006).
\bibitem{Elsaesser12}S.\ Els{\"a}sser, D.\ Wu, M.\ Dressel, and J.\ A.\ Schlueter,
Phys.\ Rev.\ B {\bf 86}, 155150 (2012);
%\bibitem{Pustogow15}
A. Pustogow, E. Zhukova, B. Gorshunov, M. Pinteri\'c, S. Tomi\'c, J. A. Schlueter, and M. Dressel,
%Low-Energy Excitations in the Quantum Spin-Liquid -(BEDT-TTF)2Cu2(CN)3
arXiv:1412.4581.

%\bibitem{Poirier12}M. Poirier, S. Parent, A. C{\^o}t\'e, K. Miyagawa, K. Kanoda, and Y. Shimizu, Phys. Rev. B {\bf 85}, 134444 (2012).

%\bibitem{Manna10} R. S. Manna, M. de Souza, A. Br{\"u}hl, J. A. Schlueter, and M. Lang, Phys. Rev. Lett. {\bf 104}, 016403 (2010); M. Lang, R. S. Manna, M. de Souza, A. Br{\"u}hl, and J. A. Schlueter, Physica C {\bf 405}, S182 (2010).

%\bibitem{Padmalekha15}K.G. Padmalekha, M. Blankenhorn, T. Ivek, L. Bogani, J.A. Schlueter, and M. Dressel, Physica B {\bf  460}, 211 (2015).





%\bibitem{Tocchio09}L. F. Tocchio, A. Parola, C. Gros,  and F. Becca, Phys. Rev. B {\bf 80}, 064419 (2009).

\bibitem{Hotta10}
C.\ Hotta, J. Phys. Soc. Japan {\bf 72} 840 (2003); Phys.\ Rev. B {\bf 82}, 241104 (2010); Crystals {\bf 2}, 1155 (2012).

\bibitem{Naka10}
M.\ Naka and S.\ Ishihara, J.\ Phys.\ Soc.\ Jpn. {\bf 79}, 063707 (2010);
J.\ Phys.\ Soc.\ Jpn. {\bf 82}, 023701 (2013).

\bibitem{Mazumdar10}
H.\ Li, R.\ T.\ Clay, and S.\ Mazumdar, J.\ Phys.: Condens.\ Matter {\bf 22}, 272201 (2010);
S.\ Dayal, R.\ Clay, H.\ Li, and S.\ Mazumdar, Phys.\ Rev.\ B {\bf 83} 245106  (2011);
R. T. Clay, S. Dayal, H. Li, and S. Mazumdar, phys. stat sol. (b) {\bf 249}, 991 (2012).


\bibitem{Gomi13}
H. Gomi, M. Ikenaga, Y. Hiragi, D. Segawa, A. Takahashi, T.J. Inagaki, and M. Aihara,
Phys. Rev. B {\bf 87}, 195126 (2013).


\bibitem{Drichko14}
N. Drichko, R. Beyer, E. Rose, M. Dressel, J. A. Schlueter, S. A. Turunova, E. I. Zhilyaeva, and R. N. Lyubovskaya, Phys.\ Rev.\ B {\bf 89}, 075133 (2014); S. Yasin, E. Rose, M. Dumm, N. Drichko, M. Dressel, J. A. Schlueter, E. I. Zhilyaeva, S. A. Turunova, and R. N. Lyubovskaya, Physics B {\bf 407}, 1689 (2012).

%%%%%%%%%%%%%%%%%%%%%% Figure
\bibitem{Nakamura09}
K. Nakamura, Y. Yoshimoto, T. Kosugi, R. Arita, and M. Imada, J. Phys. Soc. Jpn. {\bf 78}, 083710 (2009).

\bibitem{Kandpal09}
H. C. Kandpal, I. Opahle, Y.-Z. Zhang, H. O. Jeschke, and R. Valent{\'i},
Phys. Rev. Lett. {\bf 103}, 067004 (2009);
H. O. Jeschke, M. de Souza, R. Valent{\'i}, R. S. Manna, M. Lang, and J. A. Schlueter, Phys. Rev. B {\bf 85}, 035125 (2012);
J. Ferber, K. Foyevtsova, H. O. Jeschke, and R. Valent{\'i}, Phys. Rev. B {\bf 89}, 205106 (2014).

\bibitem{Oshima88}
K. Oshima, T. Mori, H. Inokuchi, H. Urayama, H. Yamochi, and G.
Saito, Phys. Rev. B {\bf 38}, 938 (1988).

\bibitem{Komatsu96}
T. Komatsu, N. Matsukawa, T. Inoue, and G. Saito, J. Phys. Soc.
Jpn. {\bf 65}, 1340 (1996).

\bibitem{McKenzie98} R.H. McKenzie, Comments Cond.\ Mat.\ {\bf 18}, 309 (1998).
%%%%%%%%%%%%%%%%%%%%%%%%%%%%






\bibitem{Abdel-Jawad10}
M.\ Abdel-Jawad, I.\ Terasaki, T.\ Sasaki, N.\ Yoneyama, N.\ Kobayashi, Y.\ Uesu, and C.\ Hotta, Phys.\ Rev.\ B {\bf 82}, 125119 (2010).

\bibitem{Pinteric14}
M. Pinteri\'c, M. \v{C}ulo, O. Milat, M. Basleti\'c, B. Korin-Hamzi\'c, E. Tafra, A. Hamzi\'c, T. Ivek, T. Peterseim, K. Miyagawa, K. Kanoda, J. A. Schlueter, M. Dressel, and S. Tomi\'c,
Phys.\ Rev.\ B {\bf 90}, 195139 (2014);
M. Pinteri\'c, T. Ivek, M. \v{C}ulo, O. Milat, M .Basleti\'c, B. Korin-Hamzi\'c, E. Tafra,
A. Hamzi\'c, M. Dressel, and S. Tomi\'c,
Physica B {\bf 460}, 202 (2015).

\bibitem{Itoh13}
K.\ Itoh, H.\ Itoh, M.\ Naka, S.\ Saito, I.\ Hosako, N.\ Yoneyama, S.\ Ishihara, T.\ Sasaki, and S. \ Iwai,
Phys.\ Rev.\ Lett.\ {\bf 110}, 106401 (2013).

\bibitem{Gorshunov05} B. P. Gorshunov, A. Volkov, I. E. Spektor, A. S. Prokhorov, A. A. Mukhin, M. Dressel, S. Uchida, and A. Loidl, Int. J. Infrared Millimeter Waves {\bf 26}, 1217 (2005).


\bibitem{Shimizu06}
Y. Shimizu, K. Miyagawa, K. Kanoda, M. Maesato, and G. Saito,
Phys. Rev. B {\bf 73}, 140407 (2006); K. Kanoda (private
communication).

\bibitem{Yamamoto12}
T. Yamamoto, K. Matsushita, Y. Nakazawa, K. Yakushi, M. Tamura,
and R. Kato (unpublished).

\bibitem{Sedlmeier12}
K.\ Sedlmeier, S.\ Els{\"a}sser, D.\ Neubauer, R.\ Beyer, D.\ Wu, T.\ Ivek, S.\ Tomi\'{c}, J.\ A.\ Schlueter, and M.\ Dressel, Phys.\ Rev.\ B {\bf 86}, 245103 (2012).

%%%%%%%%%%%%%%%%%%%%%%%%
\bibitem{SM}See Supplemental Material at http://link.aps.org./ supplemental/ for more details, a complete list and 3D motion pictures of the calculated vibrational frequencies.



%\bibitem{Hohenberg64}
%P. Hohenberg and W. Kohn, Phys. Rev. {\bf 136 B}, 864 (1964); W. Kohn and L. J. Sham,
%Phys. Rev. {\bf 140 B}, 1133 (1965).

\bibitem{Kresse93}
G. Kresse and J. Hafner, Phys. Rev. B {\bf 47}, 558 (1993) and
{\bf 48}, 13115 (1993);
G. Kresse and J. Furthm\"{u}ller, Comput. Mat. Sci. {\bf 6}, 15 (1996);
G. Kresse and J. Furthm\"{u}ller, Phys. Rev. B {\bf 54}, 11169 (1996).

\bibitem{Blochl94}
P. E. Bl\"{o}chl, Phys. Rev. B {\bf 50}, 17953 (1994).

\bibitem{Kresse99}
G. Kresse and D. Joubert, Phys. Rev. B {\bf 59}, 1758 (1999)


%%%%%%%%%%%%%%%%%%%%%%%


\bibitem{Bignozzi94}
C. A. Bignozzi, C. Chiorboli, M. T. Indelli, F. Scandola, V. Bertolasi, and G. Cilli,
J. Chem. Soc. Dalton Trans. {\bf 1994}, 2391 (1994).

\bibitem{Shatruk07}
A. Shatruk, A. Dragulescu-Andrasi, K. E. Chambers, S. A. Stoian, E. L. Bominaar, C. Achim and K. R. Dunbar,
J. Am. Chem. Soc. {\bf 129}, 6104 (2007).

\bibitem{Dumont13}
M. F. Dumont, O. N. Risset, E. S. Knowles, T. Yamamoto, D. M. Pajerowski, M. W. Meisel, and D. R. Talham,
Inorg. Chem. {\bf 52}, 4494 (2013).

\bibitem{Geiser91}
U.\ Geiser, H.\ H.\ Wang, K.\ D.\ Carlson, J.\ M.\ Williams, H.\ A.\
Charlier, J.\ E.\ Heindl, G.\ A.\ Yaconi, B.\ J.\ Love, M.\ W.\ Lathrop, J.\
E.\ Schirber, D.\ L.\ Overmyer, J.\ Q.\ Ren, and M.\ -H.\ Whangbo, Inorg.
Chem.\ {\bf 30}, 2586 (1991).


\bibitem{Foury2015}
P. Foury-Leylekian et al., to be submitted (2015)

\bibitem{Togo08}
A. Togo, F. Oba, and I. Tanaka, Phys. Rev. B {\bf 78}, 134106 (2008).

%\bibitem{Dressel04}
%M. Dressel and N. Drichko, Chem. Rev. {\bf 104}, 5689 (2004);
%D. Faltermeier, J. Barz, M. Dumm, M. Dressel, N. Drichko, B. Petrov, V. Semkin, R. Vlasova, C. M\'ezi{\`e}re, and P. Batail, Phys. Rev. B {\bf 76}, 165113 (2007).

\bibitem{Hiramatsu15}
T. Hiramatsu, Y. Yoshida, G. Saito, A. Otsuka, and H. Yamochi,
%Quantum spin liquid: design of a quantum spin liquid next to a superconducting state based on a dimer-type ET Mott insulator†
J. Mater. Chem. C {\bf 3}, 1378 (2015).

\bibitem{Dressel92}
M. Dressel, J.E. Eldridge, J. M. Williams, and  H. H. Wang,
Physica C {\bf 203}, 247 (1992).

\bibitem{Girlando00}
A. Girlando, M. Masino, G. Visentini, R. G. Della Valle, A. Brillante, and E. Venuti,
Phys. Rev. B {\bf 62}, 14476 (2000).

\bibitem{Dressel12}
M. Dressel, M. Dumm, T. Knoblauch, and M. Masino,
Crystals {\bf 2}, 528 (2012).

%%%%%%%%%%%%%%%%%%%%%%%%%%%%%%%%%%
\bibitem{remark1}
A.S. Davydov, {\it Solid State Theory} (Nauka, Moscow, 1976) (in Russian), Ch.~III;
G.D. Mahan, {\it Many Particle Physics} (Plenum, New York and London, 1981), Chs.~4 and 8.1.

\bibitem{Kuzmenko}
 E. Cappelluti, L. Benfatto, and A.B. Kuzmenko, Phys. Rev B {\bf 82}, 041402 (2010).

%\bibitem{remark2}
%A.S. Davydov, {\it Solid State Theory} (Nauka, Moscow, 1976) (in Russian), Chap.~X;
%G.D. Mahan, {\it Many Particle Physics} (Plenum, New York and London, 1981), Chap. 8.1.


\bibitem{Pouget}
P. Alemany, J-P. Pouget and E. Canadell, Phys. Rev B {\bf 85}, 195118 (2012)


%%%%%%%%%%%%%%%%%%%%%%%%%%%%%%%%%%





\end{thebibliography}

\begin{thebibliography}{99}

\bibitem[S1]{DresselGruner02}M.~Dressel and G.~Gr\"{u}ner, {\it Electrodynamics of Solids}  (Cambridge University Press, Cambridge, 2002).
\bibitem[S2]{Gorshunov05} B. P. Gorshunov, A. Volkov, I. E. Spektor, A. S. Prokhorov, A. A. Mukhin, M. Dressel, S. Uchida, and A. Loidl, Int. J. Infrared Millimeter Waves {\bf 26}, 1217 (2005).

\bibitem[S3]{Dion04}
M. Dion, H. Rydber, E. Schr\"{o}der, D. C. Langreth, and B. I. Lundqvist, Phys. Rev. Lett. {\bf 92}, 246401 (2004).
\bibitem[S4]{Roman09}
G. Rom\'{a}n-P\'{e}rez and J. M. Soler, Phys. Rev. Lett. {\bf 103}, 096102 (2009).
\bibitem[S5]{Klimes11}
J. Klime\v{s}, D. R. Bowler, and A. Michaelides, Phys. Rev. B {\bf 83}, 195131 (2011).
\bibitem[S6]{Mittendorfer11}
F. Mittendorfer, A. Garhofer, J. Redinger, J. Klime\v{s}, J. Harl, and G. Kresse, Phys. Rev. B {\bf 84} 201401 (2011).
\bibitem[S7]{Monkhorst76}
H. J. Monkhorst and J. D. Pack, Phys. Rev. B {\bf 13}, 5188 (1976).

\bibitem[S8]{Dressel04}
M. Dressel and N. Drichko, Chem. Rev. {\bf 104}, 5689 (2004).
\bibitem[S9]{Faltermeier07}
D. Faltermeier, J. Barz, M. Dumm, M. Dressel, N. Drichko, B.
Petrov, V. Semkin, R. Vlasova, C. M\'ezi{\`e}re, and P. Batail, Phys.
Rev. B {\bf 76}, 165113 (2007);
J. Merino, M. Dumm, N. Drichko, M. Dressel, and R.H. McKenzie,
Phys. Rev. Lett. {\bf 100}, 086404 (2008);
M. Dumm, D. Faltermeier, N. Drichko, M. Dressel, C. M\'ezi{\`e}re, and
P. Batail, Phys. Rev. B {\bf 79}, 195106 (2009).
\bibitem[S10]{Ferber14}
J. Ferber, K. Foyevtsova, H. O. Jeschke, and R. Valent{\'i}, Phys. Rev. B {\bf 89}, 205106 (2014).


\end{thebibliography}
\end{document}